%% file: main.tex
\documentclass[amsfonts, amssymb, amsmath, reprint, showkeys, nofootinbib]{revtex4-1}
\usepackage[english]{babel}
\usepackage[utf8]{inputenc}
\usepackage[colorinlistoftodos, color=green!40, prependcaption]{todonotes}
\usepackage{changes}
\usepackage{xcolor}
\usepackage{soul}
\usepackage{float}
\setcommentmarkup{\bfseries\color{red}-- #1 --}
\input{preamble}
\usepackage[pdftex, pdftitle={Article}, pdfauthor={Author}]{hyperref} 

\begin{document}

\title{A continuum of bright and dark pulse states in a photonic-crystal resonator}

\author{Su-Peng Yu}
\email[Correspondence email address: ]{supeng.yu@colorado.edu}
    \affiliation{Time and Frequency Division, National Institute of Standards and Technology, Boulder, CO USA}
    \affiliation{Department of Physics, University of Colorado, Boulder, CO 80309, USA}

\author{Erwan Lucas}
    \affiliation{Time and Frequency Division, National Institute of Standards and Technology, Boulder, CO USA}
    \affiliation{Department of Physics, University of Colorado, Boulder, CO 80309, USA}

\author{Jizhao Zang}
    \affiliation{Time and Frequency Division, National Institute of Standards and Technology, Boulder, CO USA}
    \affiliation{Department of Physics, University of Colorado, Boulder, CO 80309, USA}

\author{and Scott B. Papp}
\email[Correspondence email address: ]{scott.papp@nist.gov}
    \affiliation{Time and Frequency Division, National Institute of Standards and Technology, Boulder, CO USA}
    \affiliation{Department of Physics, University of Colorado, Boulder, CO 80309, USA}

\date{\today} 

\begin{abstract}

Nonlinearity is a powerful determinant of physical systems. Controlling nonlinearity leads to interesting states of matter and new applications. In optics, diverse families of continuous and discrete states arise from balance of nonlinearity and group-velocity dispersion (GVD). Moreover, the dichotomy of states with locally enhanced or diminished field intensity depends critically on the relative sign of nonlinearity and either anomalous or normal GVD. Here, we introduce a resonator with unconditionally normal GVD and a single defect mode that supports both dark, reduced-intensity states and bright, enhanced-intensity states. We access and explore this dark-to-bright pulse continuum by phase-matching for soliton generation with a photonic-crystal resonator, which mediates the competition of nonlinearity and normal GVD. These stationary temporal states are coherent frequency combs, featuring highly designable spectra and ultralow noise repetition-frequency and intensity characteristics. The dark-to-bright continuum illuminates physical roles of Kerr nonlinearity, GVD, and laser propagation in a gapped nanophotonic medium. 

\end{abstract}

\maketitle


\section{Introduction}

Complex systems generate patterns from a fundamental set of rules, which govern interactions between system components. Fractals are a good example \cite{FractalEverwhere} in which mathematical relations produce intricate patterns, depending on a set of parameters that characterize the relations. Similarly, in nonlinear optics, spatiotemporal laser patterns readily manifest in a medium, and their dynamics enable detailed, precise, and controllable tests of how light and matter interact. We focus on a ubiquitous nonlinearity of materials, the Kerr effect, \cite{Moss2013, Robson2017} which underlies fascinating behaviours in nonlinear optics, such as the formation of discrete states of patterns and pulses \cite{Lugiato1999, Barland2002}. However, the Kerr effect only represents half of the picture in the dynamics, since the frequency dependence of group-velocity dispersion (GVD or simply dispersion) typically controls what optical state forms at the balance \textcolor{black}{against} nonlinearity. Simply the sign of GVD differentiates optical states, for example in the case of anomalous GVD that balances with the Kerr effect for soliton formation. Beyond illuminating complex systems, nonlinear-optical states are being applied and optimized. Dissipative Kerr solitons in microresonators enable ultraprecise optical-frequency metrology~\cite{Drake2019} and many other functionalities~\cite{MarinPalomo2017, Suh2016, Trocha2018, Fulop2018}. More advanced device topologies promise to yield enhanced nonlinear laser sources, through coupled resonators~\cite{Xue2015, Nazemosadat2019}, dispersion engineering in nanophotoncis~\cite{Kim2017, Yu2019FP}, and inverse-design methods~\cite{CastelloLurbe2014, Vercruysse2020}. 

To understand the interplay of nonlinearity and either anomalous or normal GVD, we utilize the mean-field Lugiato-Lefever equation (LLE) \cite{Lugiato2018} for the field in a Kerr resonator. The LLE describes several states: the flat state of a sufficiently low intensity pump laser; oscillatory Turing patterns that extend over the entire resonator; and localized bright and dark solitons, which are the canonical stationary states at anomalous and normal GVD, respectively. In particular, the LLE framework for optical states subject to normal GVD (and a positive nonlinear coefficient) is relatively sparse, since the nearly unconditional imbalance of GVD and nonlinearity at constant excitation suppresses phase matching \cite{Godey2014}. Still, particular nonlinear states have been observed through fortuitous mode-structure defects that create \textcolor{black}{bands of} anomalous GVD in \textcolor{black}{an otherwise} normal-GVD resonator to seed dark-soliton formation \cite{Xue2015, Fulop2018}. 

Experiments with dark-soliton states exhibit strikingly different behaviour from its anomalous GVD counterpart, for example, the development of complex spectral modifications with pump-laser detuning \cite{Nazemosadat2019}. Moreover, effectively bright-pulse states such as the \textit{platicon} have been described through simulation \cite{Lobanov2015} and can form through multi-frequency pumping \cite{Liu2020} or the Raman effect \cite{Yao2020}. Beside interesting physics, normal GVD systems are advantageous for applications, including the relative ease of obtaining normal GVD, self-starting pulses \cite{Lobanov2015}, and focused spectral power distribution. These characteristics provide complementary functionalities to the current paradigm for laser synthesis~\cite{Spencer2018} and optical clockwork~\cite{Drake2019} with anomalous dispersion Kerr combs. It is therefore important to understand the physics underlying the emergence of normal GVD solitons, and to develop reliable methods to create these curious states.

Here, we phase-match for pattern formation in normal GVD, using a photonic-crystal resonator (PhCR) \cite{Yu2020}. The tailored point-defect in PhCR dispersion enables us to discover and assess a complete continuum of bright- and dark-pulse states under normal GVD. The PhCR defect modifies the pump-versus-loss energy balance within the resonator, providing the key that uncovers dark- and bright-pulse states in the same physical device. A PhCR is a microresonator with periodic modulation that demonstrates Bloch symmetry~\cite{PhCBook}, opening a bandgap in the resonator dispersion. We use an edge-less boundary condition-- an azimuthally uniform pattern around the resonator --to create the frequency-domain equivalent of a point-defect on a targeted azimuthal mode of the resonator \cite{McGarveyLechable2014}. By tuning a pump laser onto resonance with this point-defect mode, we phase-match for modulation instability in the normal GVD regime. Further, by designing the PhCR bandgap for specific regimes of pump laser power and detuning, we can realize both bright- and dark-pulse states of the resonator field. Our experiments explore their tuning behaviour with the pump laser and bandgap, establishing a full continuum between the bright- and dark-pulse states. Moreover, we characterize the utility of states in the dark-to-bright soliton continuum for applications through ultraprecise optical-frequency measurements.

\begin{figure}[htb]
\centering
\includegraphics[width=0.95\linewidth]{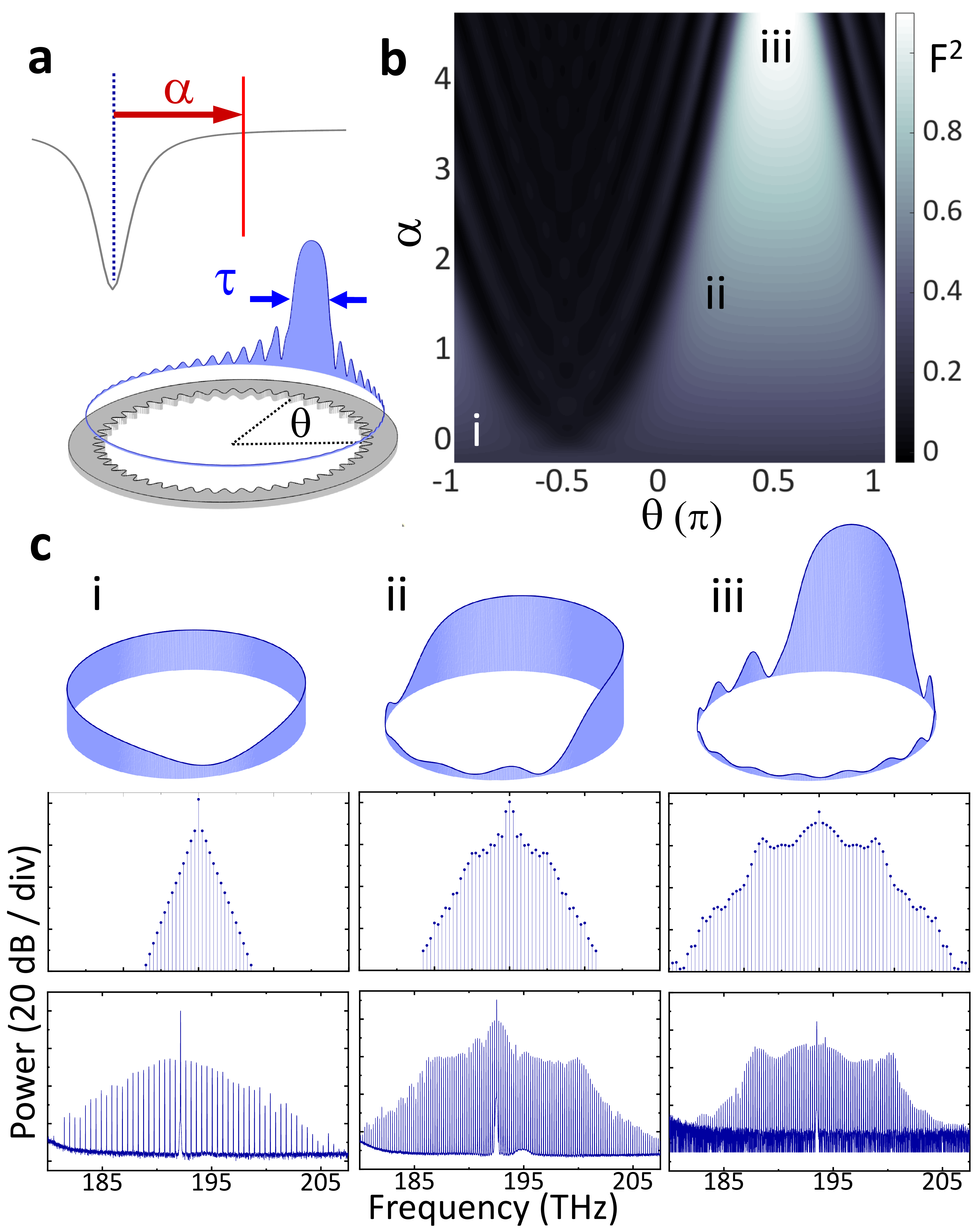}
\caption{\textbf{a} Illustration of the relation between detuning $\alpha$ and pulse width $\tau$. \textbf{b} Continuous pump detuning sweep showing \textbf{i} dark pulse, \textbf{ii} half-filled, and \textbf{iii} bright-pulse states. \textbf{c} Simulated intensities \textcolor{black}{(top)}, spectra \textcolor{black}{(middle)}, and measured spectra \textcolor{black}{(bottom)} of each state \textbf{i}--\textbf{iii}.}
\label{Fig:Intro}
\end{figure}

\section{Phenomenon}

Unifying and controlling the available states in normal-GVD Kerr resonators is an important objective; see Fig. 1. The bandgap of our PhCR devices enables a controllable frequency shift of the mode excited by the pump laser, which unconditionally satisfies phase matching for four-wave mixing. In experiments on this system, we observe spontaneous formation of optical states with spectra that suggest localized patterns. Analyzing these states highlights the curious characteristic in which the high-intensity duration $\tau$ (Fig.~\ref{Fig:Intro}\textbf{a}) varies dramatically with the laser detuning \textcolor{black}{$\alpha = \omega_\text{r} - \omega_\text{l}$}, where $\alpha>0$ for laser frequency \textcolor{black}{$\omega_\text{l}$} lower than the resonance \textcolor{black}{$\omega_\text{r}$}. Indeed, we identify these states as pulses that transition continuously from dark to bright for a prescribed tuning range of $\alpha$.  We characterize this by the \emph{intensity filling fraction} \textcolor{black}{$t_c = \tau/\tau_\text{rep}$}, \textcolor{black}{which is the pulse duration normalized to the round-trip time $\tau_\text{rep}$} or equivalently the fraction of the azimuthal angle $\theta$ occupied by higher-than-average optical intensity. Figure \ref{Fig:Intro}\textbf{b} illustrates this dark-to-bright continuum with an accurate LLE simulation for a PhCR device. The PhCR shift \textcolor{black}{modifies} the phase-matching condition and therefore what \textcolor{black}{states} can be reached in the resonator \cite{Yu2020}. Here we choose a pump-laser mode (number $\mu=0$) red-shifted by 2.0 $\kappa$ from the baseline normal GVD, where the half-linewidth of the resonator is $\kappa / 2\pi$. The plot shading indicates the continuous evolution between diminished and enhanced peak intensity, state \textbf{i} and \textbf{iii} in Fig.~\ref{Fig:Intro}\textbf{b}, respectively. The intermediate state \textbf{ii} represents a half-filled resonator.

Figure~\ref{Fig:Intro}\textbf{c, i--iii} analyze specific states in the dark-to-bright soliton continuum, making the connection between simulated spectra and pulse waveforms and our spectrum measurements of states created in normal-GVD PhCR devices. As a function of $\alpha$ starting near zero, the stationary state begins as a dark soliton~\cite{Xue2015} where a localized intensity dip exists over an otherwise flat background, i.e. the anti-pulse state \textbf{i}. As $\alpha$ increases, the anti-pulse grows in duration and develops multiple intensity minimums that extend about $\theta$ \cite{Nazemosadat2019}. State \textbf{ii} shows the dark pulse with five minimums and occupying approximately half of the $\theta$ space within the resonator. We equivalently interpret this half-filled state as a bright pulse occupying the other half of the resonator. Increasing $\alpha$ further causes the bright section to shorten temporally and increase in intensity to form the \emph{platicon state} \cite{Lobanov2015, Liu2020}, which phenomenologically describes this bright-pulse state. Indeed platicons are described by their plateau-like temporal shape and localized oscillations trailing the pulse (state \textbf{iii}). Our results provide the link between the bright- and dark-pulse states as the two extremes of a continuously tuned intensity pattern.

According to the intrinsic temporal- and spectral-domain relationships of nonlinear states, the pulse waveforms spanning the dark-to-bright pulse continuum in Fig.~\ref{Fig:Intro} exhibit identifying features. We focus on three primary features according to their spectral-domain appearance: \textit{center lobe}, \textit{wing}, and \textit{horn}. The center lobe refers to the high-power modes near the pump laser, and \textcolor{black}{its bandwidth inversely represents the temporal size of a bright or dark localized pulse.} Indeed, both dark- (Fig.~\ref{Fig:Intro}\textbf{i}) and bright-pulse states (Fig.~\ref{Fig:Intro}\textbf{iii}) prominently feature a center lobe, and the prototype dark soliton is almost entirely composed of it. The more exotic wing feature develops outside the bandwidth of the center-lobe at larger $\alpha$. It represents the deviation of the temporal pattern from a pulse and is most visible when the pattern is temporally extended (Fig.~\ref{Fig:Intro}\textbf{ii}), reminiscent of the spectrum of a bandwidth-limited square wave. Indeed, the power spectrum of the half-filled state is primarily composed of the $\mu=\pm 1$ modes and a $1/\mu^2$ asymptotic envelope forms the wing. Moreover, the half-filled state demarcates the condition of center-lobe broadening and temporal pulse compression either to a dark or bright state with any change to $\alpha$. This behavior stands apart from anomalous dispersion Kerr solitons~\cite{Guo2017, Godey2014} in which increasing $\alpha$ leads to a monotonic increase in the soliton bandwidth while approximately maintaining the power-per-line near the pump \cite{Lucas2017}. The horn feature refers to the heightened spectral power on the edges of the spectral bandwidth (Fig.~\ref{Fig:Intro}\textbf{iii}). It represents the rapid oscillation trailing the pulse, a normal GVD correspondence of the dispersive waves \cite{Brasch2016cherenkov}. \textcolor{black}{These features help us extract waveform information from the optical spectra in the following sections.} 


\section{Mechanism} 

We develop a theoretical framework for the pulse-duration-tuning behavior throughout the dark- to bright-pulse continuum. Specifically, we establish the relation between the intensity filling ratio $t_c$ and the laser detuning $\alpha$. Normal GVD waveforms consist of two intensity levels, which depend on $\alpha$ and the pump $ F $. The two levels are connected by switching fronts~\cite{Coen1999}. Here, we show that the PhCR disturbs these levels through an \emph{effective} pump contribution $F'$, which introduces an additional energy exchange between the two levels that determines $t_c$.

\begin{figure}[t!]
	\centering
	\includegraphics[width=\linewidth]{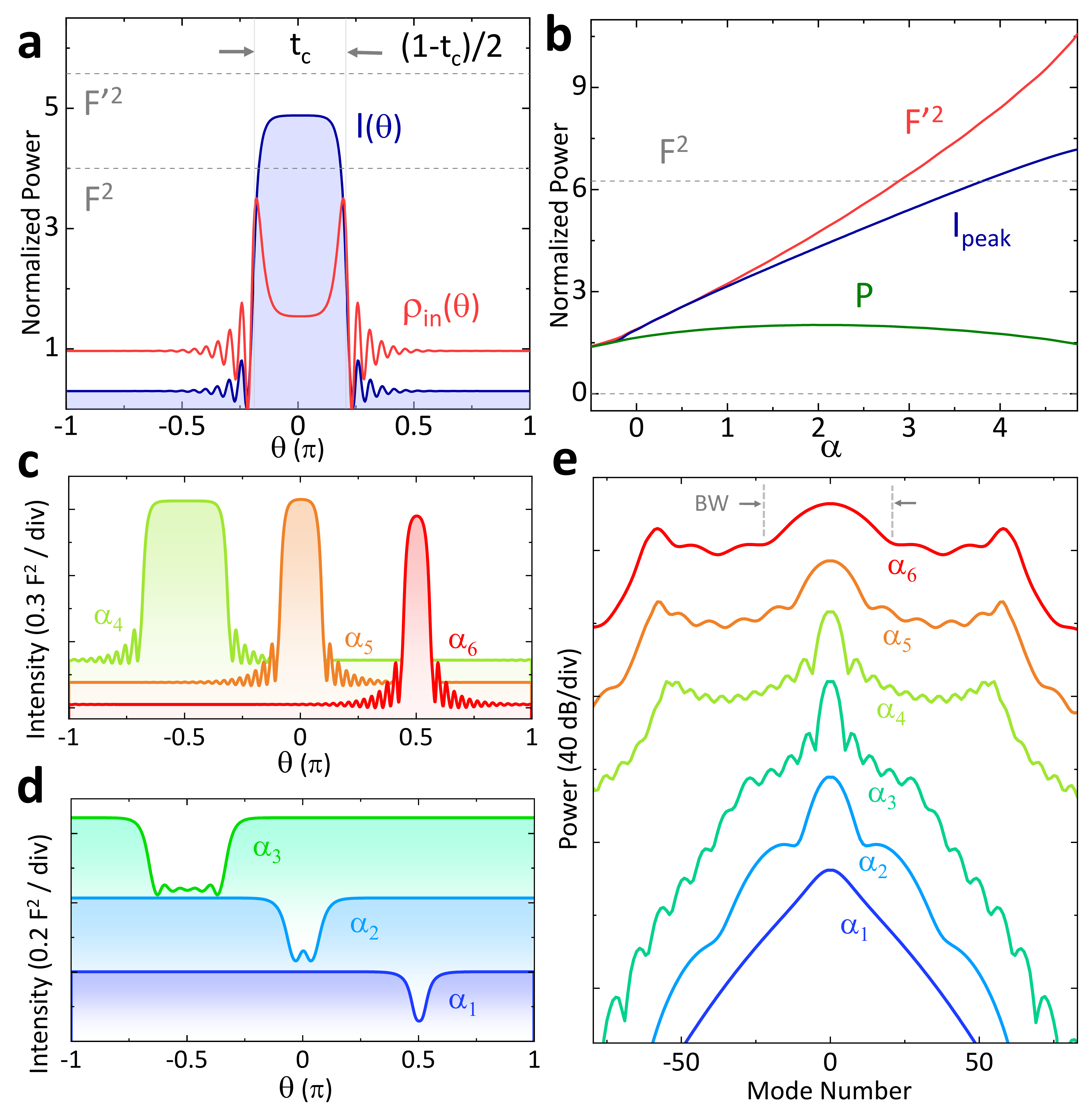}
	\caption{
		\textbf{a} Intensity profile $I(\theta)$ of a waveform with bright-pulse fraction $t_c$. $F^2<I(\theta)<F'^2$ for the high-intensity level. The pump energy in-flow $\rho_\text{in}(\theta)$ is also shown. \textbf{b} Calculated parameters $F'^2$, $I_{peak}$, and $P$ in the LLE. Note that $I_{peak}$ > $F^2$ at high $\alpha$.
		\textbf{c} and \textbf{d} plot the waveform on the red- and blue-detuned ranges of an $\alpha$ sweep ($\alpha_1<\alpha_2 ... <\alpha_6$), showing the increase of dark-pulse duration and the decrease of bright-pulse duration with $\alpha$. The corresponding spectra are shown in \textbf{e}. Note the behaviour of the center lobe bandwidth BW marked in gray.
	}
	\label{Fig:Theory}
\end{figure}


We begin exploring the relation $t_c(\alpha)$ by characterizing the waveform under normal GVD, using the pump-shifted LLE (hereafter PS-LLE) \cite{Yu2020} that accurately describes the field of the PhCR:
\begin{equation}\label{Eq:PSLLE}
    \partial_t \psi = -(1+i\alpha)\psi -\frac{i\beta}{2}\partial_\theta^2 \psi + i |\psi|^2 \psi+F + i \epsilon (\Bar{\psi}-\psi),
\end{equation}
\noindent where $\Bar{\psi}$ denotes the average field over $\theta$.
For time-stationary solutions, $     \partial_t \psi = 0 $, the interaction between dispersion and nonlinearity manifests in the imaginary part of this equation. \textcolor{black}{This yields} in absence of the PhC term (assuming a real-values pulse $ \psi $ for simplicity): $\frac{-\beta}{2}\partial_\theta^2\psi/\psi+|\psi|^2 \simeq \alpha$. For example, in anomalous GVD ($\beta<0$), the peak of a pulse shows negative curvature and positive Kerr shift. Partial cancellation or \textit{balance} between the two enables sharp waveforms like the Kerr soliton. Our system is in normal GVD ($\beta>0$), leading to \textit{competition} between dispersion and nonlinearity, which sum to a particular $\alpha$ to phase-match to the pump laser. This leads to a mutually \textcolor{black}{exclusive} relation between local intensity and curvature-- where the intensity is high, the waveform is flat --leading to the flat-top waveform with switching edges in-between shown in Fig.~\ref{Fig:Theory}\textbf{a}. $t_c$ corresponds to the fraction of the high-intensity level.

This two-level waveform already exists in the normal-GVD regime of conventional resonators~\cite{ParraRivas2016}. The waveform corresponds to the flat-amplitude levels of the bi-stable, continuous-wave (CW)  resonator field~\cite{Coen1999}, controlled by $F$ and $\alpha$ through solutions to $F^2 = (1+(|h|^2-\alpha)^2) |h|^2$ \cite{Godey2014}, where formally $h = F/(1+i(\alpha-|h|^2))$ is the field at each level. These levels are stationary in time because the loss $\rho_\text{loss}$ and pump power in-flow $\rho_\text{in}$ per unit $\theta$ balance
\begin{align} \label{Eq:EnergyFlow}
	\rho_\text{loss}(\theta) &= \kappa \cdot I(\theta) = |\psi(\theta)|^2 \\
	\rho_\text{in}(\theta) &= F \cdot \mathbb{R}e(\psi(\theta))
\end{align}
where $I(\theta)$ is the intensity, $\kappa = 1$ is the normalized loss rate; see \textit{Supplementary} Sec.~\ref{S:RhoDerive}. In a conventional, normal-GVD resonator, the balance $\rho_\text{in} = \rho_\text{loss}$ is satisfied \textit{locally} for all $\theta$ at the bi-stability levels. Therefore, switching edges can translate independently about $\theta$ without perturbing the input-output energy flow, although weak oscillating tails of the edges can trap some waveforms in a specific configuration~\cite{ParraRivas2016}. 


The mode structure of a PhCR perturbs the energy balance of the two levels, leading to novel nonlinear dynamics. We identify the impact of the PhCR frequency shift term in Eq.~\eqref{Eq:PSLLE} by casting it in a conventional LLE with \emph{effective} pump parameters $F'$, $\alpha'$:
\begin{align}
	F' &= |F + i \epsilon \, \Bar{\psi}| \\
	\alpha' &= \alpha + \epsilon
\end{align}%
where the corresponding field amplitudes within each intensity level (neglecting the effect of curvature) are written as
\begin{equation}
 h = \frac{F'}{1+i(\alpha'-|h|^2)},
\end{equation}%
where $|h|^2$ approximates the high- and low-level intensities at $\theta = 0$ and $\pi$ in Fig.~\ref{Fig:Theory}\textbf{a}. With $h$ determined by $F'$ but $\rho_\text{in}$ still dependent on the physical pump $F$, $\rho_\text{in}$ and $\rho_\text{loss}$ are no longer equal (these rates have the same sign by our definition). In the PhCR case, $h$ at the high-intensity level is driven to be larger by the effective $F'$. This is the mechanism that creates a larger peak intensity, forming a bright pulse.

Figure~\ref{Fig:Theory}\textbf{a} shows $\rho_\text{in}(\theta)$ and $\rho_\text{loss}(\theta) = \kappa \, I(\theta)$, calculated using the PS-LLE. The high-intensity level exhibits a deficit of energy, $I>\rho_\text{in}$, while the low-intensity level shows an energy surplus, $I<\rho_\text{in}$. When combined, the two levels maintain energy conservation.
The surplus or deficit arises from the difference in the relations $I(\theta) \propto |h|^2$ and $\rho_\text{in} \propto |h|^1$, which increase with $\alpha$ and $F'$. Importantly, both $F' > F$ and high-level intensity $I(0) = |h|^2 > F^2$ can be reached at large detuning.
Figure~\ref{Fig:Theory}\textbf{b} shows the behavior of $F'$ and $I_\text{peak}$ for the same parameters as Fig.~\ref{Fig:Intro}\textbf{b}, where the peak intensity $I_\text{peak}$ in the PS-LLE corresponds to the high-level $I(0)$. We see both parameters surpass $F^2$ at large $\alpha$. To maintain the $I_\text{peak} > F^2$ intensity with the limited physical pump power $F^2$ available to the system, the pulse duration reduces as its intensity increases. This manifests as reducing $t_c$ (hence reducing the energy deficit) and also regularizing the increase of $F'$ through the $\Bar{\psi} \simeq h \cdot t_c$ dependence. As result, the total energy inflow P $ = \oint \rho_\text{in}(\theta) d \theta$ in Fig.~\ref{Fig:Theory}\textbf{b} remains relatively constant in contrast to the increase of $F'$ or $I_\text{peak}$. This \textit{energy balance} links $t_c$ to the peak field $h$, and thus to $\alpha$.

We calculate $t_c$ through the energy balance by integrating all energy flow within the resonator. We obtain the energy conservation condition by multiplying Eq.~\eqref{Eq:PSLLE} by $\psi^*$, followed by integrating the terms over $\theta$:
\begin{equation}
\begin{split}
    (1+i\alpha)\oint |\psi|^2 d\theta = &\frac{i\beta}{2} \oint |\partial_\theta \psi|^2 d\theta + i \oint |\psi|^4 d\theta \\
    & +F\Bar{\psi^*} + i \epsilon \, (|\Bar{\psi}|^2-\oint|\psi|^2 d\theta)
    \end{split}
\end{equation}
\noindent where the second-derivative in $\theta$ term is integrated by part. Taking the real part of this form, we obtain the energy-balance equation:
\begin{equation}
    \oint |\psi|^2 d\theta = \oint F \cdot \mathbb{R}e(\psi) d\theta
\end{equation}
\noindent where we identify the terms on the two sides corresponding to $ \oint I(\theta) d\theta = \oint \rho_\text{in}(\theta) d\theta$. Expressing this form approximately in terms of the fields at $\theta=0, \pi$ and $t_c$, we get
\begin{equation}
    t_c \cdot I(0) + (1-t_c) \cdot I(\pi) = t_c \cdot \rho_\text{in}(0) + (1-t_c) \cdot \rho_\text{in}(\pi),
\end{equation}
\noindent which we rearrange to $t_c = \frac{\rho_\text{in}(\pi)-I(\pi)}{\rho_\text{in}(\pi)-\rho_\text{in}(0)+I(0)-I(\pi)}$. This form indicates how $t_c$ depends explicitly on the intensities of the two levels, and therefore implicitly on $\alpha$.


Figures \ref{Fig:Theory}\textbf{c} and \ref{Fig:Theory}\textbf{d}  present time-domain PS-LLE solutions across the continuum \textcolor{black}{at $\beta=5.2\times 10^{-3}$} , describing the soliton as we vary $\alpha$ to access both dark and bright pulses. We tune $\alpha$ from a setting $\alpha_1$ that yields the dark pulse, to the longer-duration dark pulse $\alpha_3$, across the half-filled state to a setting that yields the bright pulse $\alpha_4$, then to shortening bright pulse $\alpha_6$. The PS-LLE calculations confirm the monotonic tuning of $t_c$ with $\alpha$ described in the mechanism. Fig.~\ref{Fig:Theory}\textbf{e} shows the corresponding spectra for $\alpha_1$ through $\alpha_6$. We highlight how the center-lobe bandwidth BW is governed by $t_c$. The monotonic increase of $t_c$ manifests as a lengthening dark pulse ($\alpha_1$ through $\alpha_3$) and reducing BW, but as a shortening bright pulse ($\alpha_4$ through $\alpha_6$) and increasing BW with $\alpha$. See \textit{Supplementary} Sec.~\ref{S:BWtcDerive} for details. The tuning of the center-lobe bandwidth will be the parameter we measure in our experiments.



\begin{figure}[t]
\centering
\includegraphics[width=1.0\linewidth]{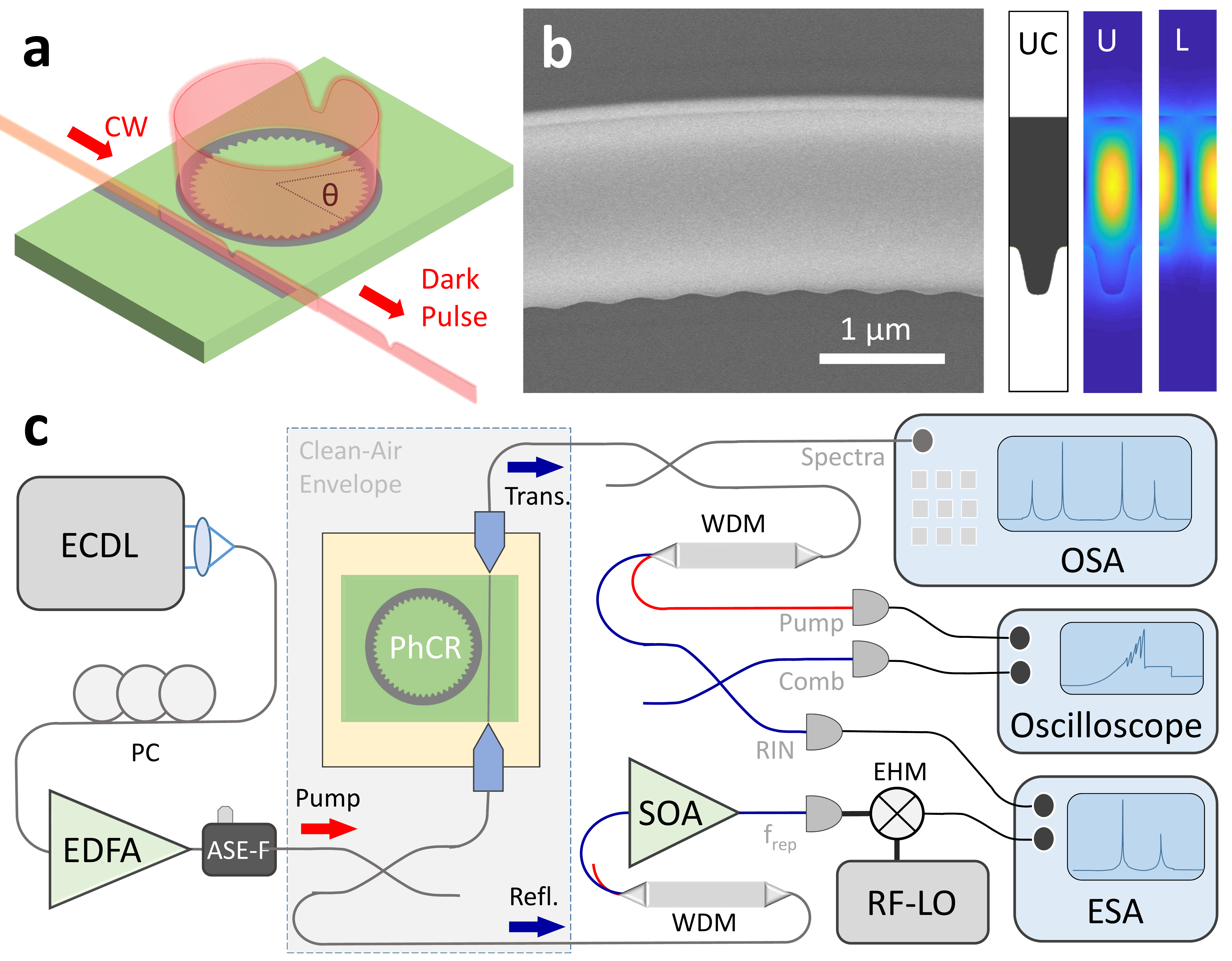}
\caption{%
	\textbf{a} Illustration of dark pulse generation in the PhCR.
	\textbf{b} Electron microscopy image showing a section of the PhCR. The unit cell (UC) is shown on the right with exaggerated modulation for clarity. The right-most panels show simulated electric field distributions for the (U) upper and (L) lower modes.
	\textbf{c} Diagram showing the optical testing setup.}
\label{Fig:System}
\end{figure}

\section{Apparatus and Procedures} 

We fabricate normal-GVD PhCR devices to explore the dark-to-bright pulse continuum. Our objectives are to: create devices with the GVD and $\epsilon$ settings that coincide with our theoretical predictions; energize the devices with a range of $\alpha$ settings to create optical states; and identify the spectral characteristics of these states. PhCRs are ring resonators with a sinusoidal modulation of the inner edge. The modulation amplitude determines the frequency shift $\epsilon$ of one mode. We couple the PhCR devices evanscently with a waveguide on the chip, illustrated in Fig.~\ref{Fig:System}\textbf{a}. We anticipate that normal-GVD solitons arise spontaneously from instability of the flat state~\cite{Lobanov2015}, and the outcoupled laser pulse forms a frequency comb, which we characterize through spectral-domain measurements.

We nanofabricate PhCRs with the tantalum pentoxide (\ce{Ta2O5}, hereafter tantala) material platform~\cite{Jung2019}. Our process begins with a 570 nm thick ion-beam sputtered film of tantala on a  3~$\mu$m thick oxidized silicon wafer. We define the pattern for PhCRs and their waveguides, using electron beam lithography, and a fluorine inductively coupled plasma reactive-ion etch (ICP-RIE) transfers the pattern to the tantala layer. We separate the wafer into chips each with several PhCR devices, using a deep Si RIE. The PhCR coupling waveguides extend to the chip edges, enabling pump laser insertion to the chip at $\sim$5 dB loss per facet. Figure~\ref{Fig:System}\textbf{b} shows a section of a PhCR with a radius of 22.5~\micro\meter, obtained with a scanning electron microscope. This PhCR unit cell indicates the amplitude and period of the modulation that controls $\epsilon$ and the azimuthal mode number, respectively. Specifically, one programmed azimuthal mode is frequency shifted to higher and lower frequency resonances, separated by a photonic bandgap. By tuning the pump laser onto resonance of the lower-frequency mode (hereafter the pump mode), we adjust the settings of normal GVD, $\epsilon$, and $\alpha$ to coincide with our detailed theoretical modeling. 


\begin{figure*}[t!]
\centering
\includegraphics[width=1.0\linewidth]{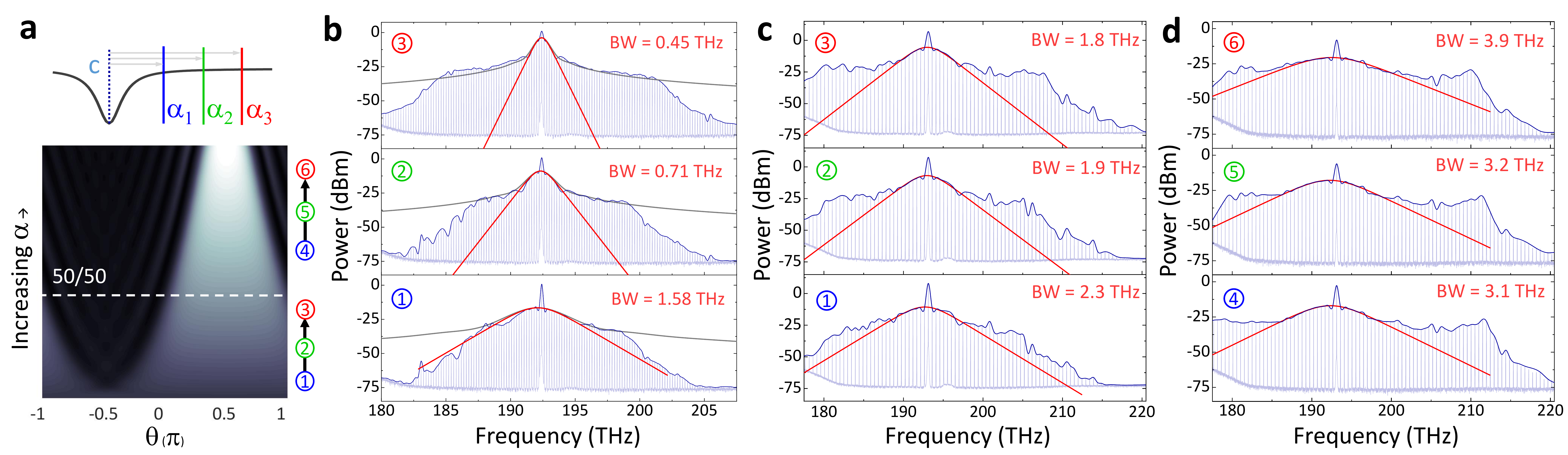}
\caption{%
	\textbf{a} Illustration of a detuning sweeps toward (1 $\rightarrow$ 3) and away from (4 $\rightarrow$ 6) the half-filled state (white dashed line), where both sweeps are from high to low frequencies ($\alpha_1 \rightarrow \alpha_3$).
	\textbf{b} Measured spectra in a FSR = 200~GHz PhCR showing the center lobe bandwidth (BW, fitted red trace) decreasing with increasing detuning. The spectral envelope of a square wave (\textcolor{gray}{\textbf{-}}) is plotted for comparison.
	\textbf{c} and \textbf{d} show spectra from nominally identical FSR = 500~GHz rings, but with peak-to-peak PhC amplitude of 8~nm and 12~nm, resulting in $\epsilon = $ 4.5 and 5.9 $\kappa$, respectively. This places soliton behaviour of the two PhCRs on opposite sides of the half-filled state and an opposing BW dependence on $\alpha$. Moreover, prominent horn features appear in \textbf{d}.
}
\label{Fig:Spectra}
\end{figure*}


Figure~\ref{Fig:System}\textbf{c} presents our test system and procedures. The pump laser is a tunable external-cavity diode laser (ECDL), and we adjust polarization by straining single-mode fiber. We amplify the pump laser with an erbium-doped fiber amplifier (EDFA), followed by a filter that suppresses amplified spontaneous emission. We mount chips-under-test to a thermally stable platform, and we align lensed fibers to the chip for input and output. We carefully design PhCRs so that the pump mode falls within the 1550 nm wavelength range, which is convenient for commercial laser components. In our experiments, the primary observable is the soliton spectrum, which we measure with an optical-spectrum analyzer (OSA) in both transmission and reflection from the chip. Therefore, we use a fiber coupler to access the reflection port and a wavelength-division multiplexer (WDM) to spectrally separate the pump laser from the transmission port. Assessing the soliton's noise characteristics is also important, since relatively low noise is an expected property of all the modelocked states across the dark-to-bright continuum. We photodetect the entire soliton, except the pump laser, and record the relative intensity noise with an electronic spectrum analyzer (ESA). Additionally, we use a $\sim150$ GHz bandwidth modified uni-traveling carrier photodetector~\cite{Jesse2018} and an ESA to record the outcoming pattern repetition frequency of suitable PhCR devices.

\section{Exploring the dark-to-bright pulse continuum}

We search for dark- and bright-pulse states in normal GVD PhCRs with parameter settings derived from our theoretical model. Figure~\ref{Fig:Spectra}\textbf{a} indicates the detuning dependence of the transition from dark-to-bright pulses with respect to the half-filled state. In a set of experiments examining both sides of the continuum, we systematically vary $\epsilon$ with discrete PhCR devices, and we vary $\alpha$ according to the sequences \textbf{1} $\rightarrow$ \textbf{2} $\rightarrow$ \textbf{3} for dark pulses or \textbf{4}$\rightarrow$ \textbf{5} $\rightarrow$ \textbf{6} for bright pulses in Fig. \ref{Fig:Spectra}\textbf{a}. For each setting of $\alpha$, we record the state's optical spectrum, and we directly analyze the center lobe, wing, and horn spectral signatures with respect to our theoretical predictions, identifying the dark- and bright-soliton pulse shapes in normal GVD. In particular, we identify the reversal of spectral bandwidth tuning behavior in response to the setting of $\alpha$ on either side of the continuum; see Fig.~\ref{Fig:Spectra}\textbf{a}. 




Figure~\ref{Fig:Spectra}\textbf{b} presents spectrum measurements of dark pulses. Specifically, we observe a decrease in the center-lobe bandwidth as a function of increasing $\alpha$. By fitting the center-lobe portion of the spectrum to a model proportional to $\sech^2(\frac{\nu-\nu_0}{\rm BW})$ where $\nu$ is optical frequency, $\nu_0$ is the center of the spectrum, and $\rm BW$ is the bandwidth, we directly characterize the center lobe. Indeed, the center-lobe bandwidth is linked to the filling fraction $t_c$ by the expression $\text{BW} = \sqrt{3} \, \text{FSR}/\pi\cdot (1-t_c)^{-1}$, where $\rm FSR$ is the free-spectral range; see \textit{Supplementary} Sec.~\ref{S:BWtcDerive}. We observe a reduction in center-lobe bandwidth from 1.58 to 0.45 THz as we increase $\alpha$.  In temporal units, the measurements in Fig.~\ref{Fig:Spectra}\textbf{b} indicate that $t_c$ varies from 0.07 to 0.25 with increasing $\alpha$, a range in agreement to our PS-LLE simulations. This data indicates a distinction of dark solitons in comparison to anomalous GVD bright solitons, which exhibit the opposite behavior with $\alpha$. Furthermore, the dark-soliton pulses develop the wing feature outside the bandwidth of the center lobe as we increase $\alpha$ and the localized dark pulse expands into the square-wave-like pattern of the half-filled state. To highlight this behavior, we overlay in Fig.~\ref{Fig:Spectra}\textbf{b} the spectral envelope of a square wave with the same bandwidth of the center-lobe. The characteristic $1/\mu^2$ roll-off of the square wave reproduces the envelope of the wing feature until the bandwidth limit set by the PhCR GVD. We note that both the center lobe and wing features behave according to our theoretical prediction in Fig.~\ref{Fig:Theory}\textbf{e}.




Figures~\ref{Fig:Spectra}\textbf{c} and \textbf{d} explore spectrum measurements in PhCRs designed to host dark and bright pulses, respectively. We use PhCRs with 500 GHz FSR and settings of $\epsilon= 4.5 \, \kappa$ (Fig.\ref{Fig:Spectra}\textbf{c}) to realize dark pulses and $\epsilon= 5.9 \, \kappa$ (Fig.~\ref{Fig:Spectra}\textbf{d}) to realize bright pulses. In an experiment, we vary $\alpha$ according to the sequences \textbf{1} $\rightarrow$ \textbf{2} $\rightarrow$ \textbf{3} for dark pulses or \textbf{4}$\rightarrow$ \textbf{5} $\rightarrow$ \textbf{6} for bright pulses by monotonically tuning the pump laser frequency. By varying $\alpha$, we explore both regimes of the dark-to-bright continuum with the half-filled state as intermiediate between them. Fitting the center lobe in both these regimes shows the characteristic increase in bandwidth as we vary $\alpha$ across the normal-GVD soliton continuum. Moreover, from our theoretical predictions in Fig.~\ref{Fig:Theory}\textbf{e}, we anticipate that these bright solitons (Fig.~\ref{Fig:Spectra}\textbf{d}) will exhibit a significant horn feature, which is the analog of more well-known dispersive waves in the anomalous-GVD regime. Our measurements demonstrate the horn feature with an $\sim$ 5 dB spectral enhancement near the PhCR GVD bandwidth limit of the soliton. Similar to the dispersive waves, the horn elevates the comb power above the center-lobe envelop, leading to the observed plateau-like spectral profile characteristic of these states. The set of normal-GVD soliton spectrum measurements in Fig.~\ref{Fig:Spectra}, obtained by tuning $\alpha$, presents a comprehensive test of the dark-to-bright continuum.




\section{Frequency-comb sources from the dark-to-bright pulse continuum}

\begin{figure}[h]
\centering
\includegraphics[width=1.0\linewidth]{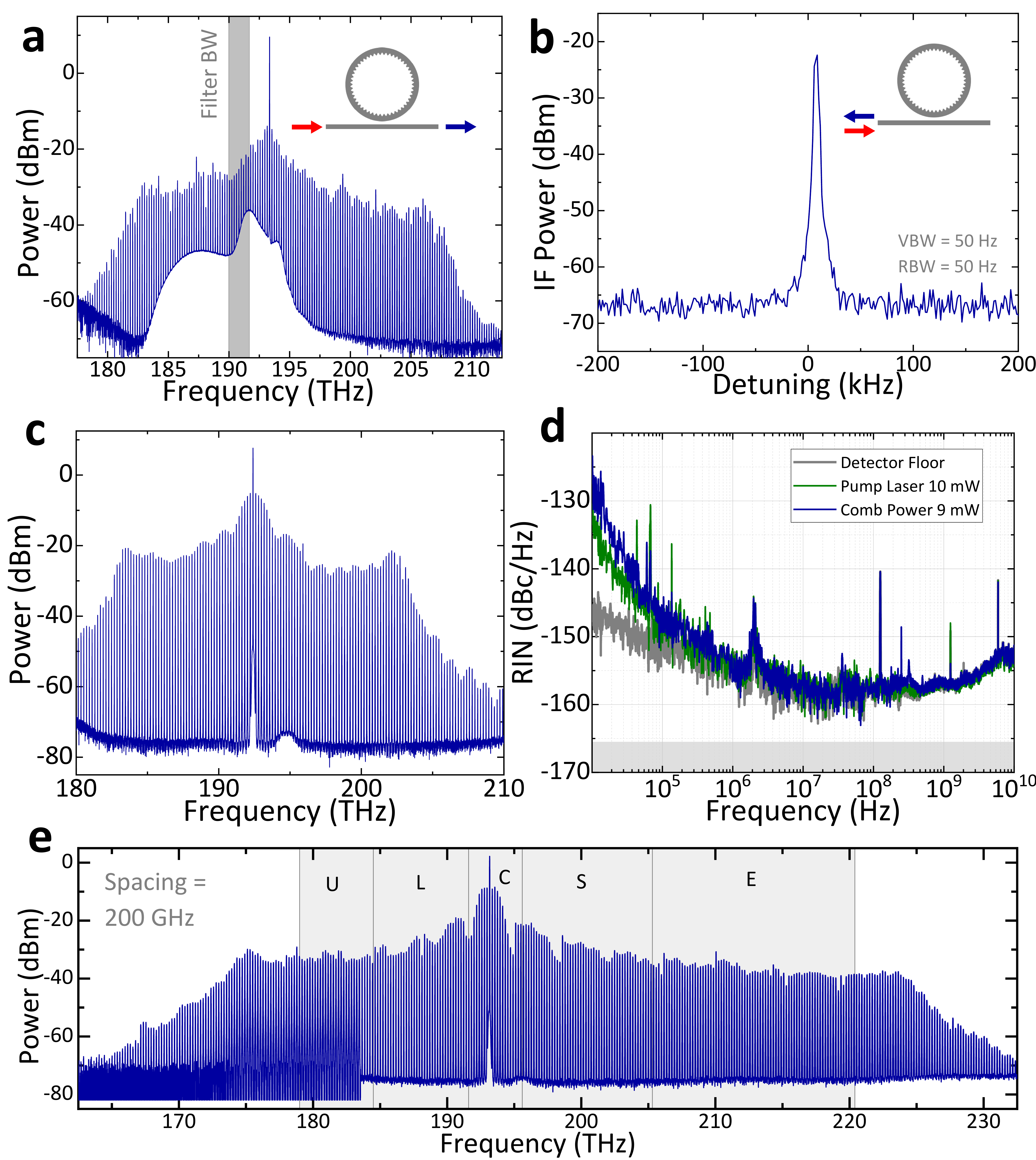}
\caption{%
	\textbf{a} Optical spectrum monitoring at the transmission port, and
	\textbf{b} down-mixed electronic repetition rate beatnote measured at the reflection port.
	\textbf{c} Optical spectrum and
	\textbf{d} corresponding measured relative intensity noise on the comb power.
	\textbf{e} Demonstration of a 200~GHz repetition rate comb covering Telecom U through S bands.
}
\label{Fig:Coherence}
\end{figure}

We anticipate that nearly any state of the dark-to-bright pulse continuum will yield a useful frequency-comb source. Moreover, the normal-GVD regime of Kerr frequency combs presents unique opportunities in terms of comb lasers with designable spectral coverage, relatively constant comb-mode power distribution, and high conversion efficiency of the pump laser to the integrated comb power. Figure \ref{Fig:Coherence} presents examples of spectral design and noise measurements with a 200 GHz FSR PhCR. 


The concept of a frequency comb is generalized from the particle-like Kerr soliton to time-stationary patterns in a resonator with a single repetition frequency. But naturally the repetition frequency and comb power vary with the parameters of the PS-LLE. Moreover, the mode-frequency splitting of our PhCRs arises from a coupling of forward and backward propagation direction, and we observe that the pulse propagation direction with respect to the pump laser primarily depends on the relative setting of $\alpha$ and $\epsilon$. The noise measurements we present here explore pulse propagation in both directions.

To characterize the repetition frequency, we operate a PhCR to generate a soliton pulse train reverse to the pump laser. For this experiment, we monitor a fraction of the comb spectrum through the transmission port; see Fig. \ref{Fig:Coherence}a. This signal results from a reflection of the backward-propagating comb to the forward transmission port. To measure the repetition frequency, we couple a portion of the comb power to a $\sim150$ GHz bandwidth, 0.2 A/W responsivity photodetector \cite{Jesse2018}. We use an optical circulator and a 2 THz bandwidth optical filter prior to photodetection to reduce photocurrent from the pump laser. We extract the 200 GHz photocurrent signal from the photodectector chip with a microwave probe, and we use a fourteenth-order harmonic mixer driven by a 13.92 GHz signal to down-convert the repetition frequency. Figure \ref{Fig:Coherence}\textbf{b} shows the repetition frequency at an intermediate 371.6 MHz frequency. The high signal to noise ratio of the repetition frequency is consistent with a low-noise frequency comb of equidistant modes operating in the soliton regime.


In a second characterization experiment, we measure the relative intensity noise (RIN), which is a critical characteristic for example in applications that the comb modes are encoded with information. Here, we operate a PhCR to generate a soliton pulse train in the forward direction with respect to the pump laser. The optical spectrum of a forward-emitting comb state in this measurement is shown in Fig.~\ref{Fig:Coherence}\textbf{c}. We separate the comb power from the transmitted pump power, using a wavelength-filtering element prior to photodetection. The photodiode has 12~GHz nominal bandwidth and 0.8 A/W responsivity to measure RIN, and we deliver the 9~mW total comb power to the detector without amplification. Figure~\ref{Fig:Coherence}\textbf{d} shows the (RIN) on the photdetected signal. The detector noise and the RIN of the pump laser are approximately at the same power level as the comb. The RIN level ranges from $-130$~dBc/Hz at 10~kHz to $-160$~dBc/Hz at higher frequencies, currently limited by the detector noise floor. 



Finally, we demonstrate how normal-GVD soliton combs in PhCRs may be used in the future. Frequency-comb lasers are revolutionizing optical communication systems, which require dense carrier grids in for example the 1300 nm and 1550 nm wavelength bands. Still, universal laser sources based on scalable photonics technology do not exist, primarily due to physical limitations of laser gain. Soliton microcombs are recognized as a promising technology for this application \cite{MarinPalomo2017, Fulop2018}, but especially in the anomalous GVD regime there has been no demonstration of a microcomb that can support multiple wavelength bands. Here, we demonstrate a normal-GVD PhCR with suitable properties to generate a broadband comb laser with relatively constant spectral envelope and a dense 200 GHz mode spacing. Figure ~\ref{Fig:Coherence}\textbf{e} shows the comb-laser spectrum, which spans the standardized telecommunication bands denoted U, L, C, S, E and a portion of the O band. Such a spectral coverage of 50 THz exceeds what is possible with either fiber-based solid-state gain materials or semiconductor gain materials, highlighting the uniqueness of microcomb technology. Moreover, our PhCR soliton microcomb laser offers high conversion efficiency from the pump laser to the comb modes. Efficiency is a critical metric in for example hyperscale data centers where the demands of ever-increasing internet traffic and services causes massive energy consumption. More efficient laser sources, especially comb lasers, are one of the most important technology areas \cite{Cheng2018}. Specifically, we characterize the comb conversion efficiency $\eta = P_\text{comb}/F^2$ from the input pump power $F^2$, and we predict that $\eta \approx 25\%$ is attainable based on modeling with the PS-LLE. In our experiments, we obtain a conversion efficiency as high as 21$\%$ in which a PhCR converts a 33 mW pump laser to soliton microcomb with 200 GHz mode spacing and 7 mW mode-integrated power that spans the optical frequency range from 180 THz to 210 THz. This information highlights the importance of normal-GVD PhCR soliton microcombs in the technology frontier.


\section{Conclusion}

We have presented a new regime of nonlinearity in which both bright and dark pulse states are stable in the same physical resonator. We control the balance of nonlinearity and loss to phase-match four-wave mixing in a normal-GVD PhCR by adjusting a mode bandgap for the pump laser. Moreover, these pulse states arise spontaneously from a CW-laser flat background, according to phase matching with the PhCR. Laser detuning is intrinsically linked to the intensity-filling fraction of a pulse state, and a pulse with an intensity dip can continuously evolve to a localized bright pulse. Indeed, at the center of the dark-to-bright continuum is the half-filled state, which represents the pulse transition edge between high and low intensity levels. Both the dark and bright pulse states manifest as a frequency comb with fingerprint spectral features, which we analyze for comparison with our detailed numerical models. In particular, we expect and observe a striking inversion of pulse-bandwidth tuning with laser detuning centered on the half-filled state that highlights the fundamental difference in the phase-matching with normal and anomalous GVD. This type of microcomb laser is a versatile and efficient multi-wavelength source with high spectral coherence, which enables various signaling and sensing applications.  

\begin{acknowledgements}
We thank Jennifer Black and Travis Briles for carefully reading the paper. This research is supported by the Defense Advanced Research Projects Agency PIPES program and NIST. EL acknowledges support from the Swiss National Science Foundation (SNSF).
\end{acknowledgements}

\vspace{5pt}
\noindent \textbf{Author Contributions} S.-P.Y. contributed in the conception, design and fabrication, and theoretical analysis; S.-P.Y., E.L., and J.Z. performed the optical and radio-frequency measurements. S.B.P. contributed to the theoretical understanding and supervised the findings of this work; All authors provided feedback and helped shape the research, analysis and manuscript.

\vspace{5pt}
\noindent \textbf{Competing Interests} The authors declare that they have no competing financial interests.

\vspace{5pt}
\noindent \textbf{Supplementary Information} is available for this paper.



\bibliographystyle{ieeetr}
\bibliography{DarkSoliton}

\clearpage

\setcounter{figure}{0} 
\setcounter{section}{0}
\setcounter{equation}{0}

\onecolumngrid
{
\centering{
\Large
\textbf{Supplementary Information for: A continuum of bright- and dark-pulse states in photonic-crystal resonators}
\vspace{12pt}

Su-Peng Yu, Erwan Lucas, Jizhao Zang, and Scott B. Papp$^\ast$
\vspace{5pt}

\small
Time and Frequency Division, National Institute of Standards and Technology, Boulder, CO 80305, USA

Department of Physics, University of Colorado, Boulder, CO 80309, USA

 $^\ast$Correspondence email address: scott.papp@nist.gov

}
}

\vspace{25pt}
\normalsize
\twocolumngrid

\section{Phase-Matching in Normal Dispersion}

\begin{figure}[htb]
\centering
\includegraphics[width=0.95\linewidth]{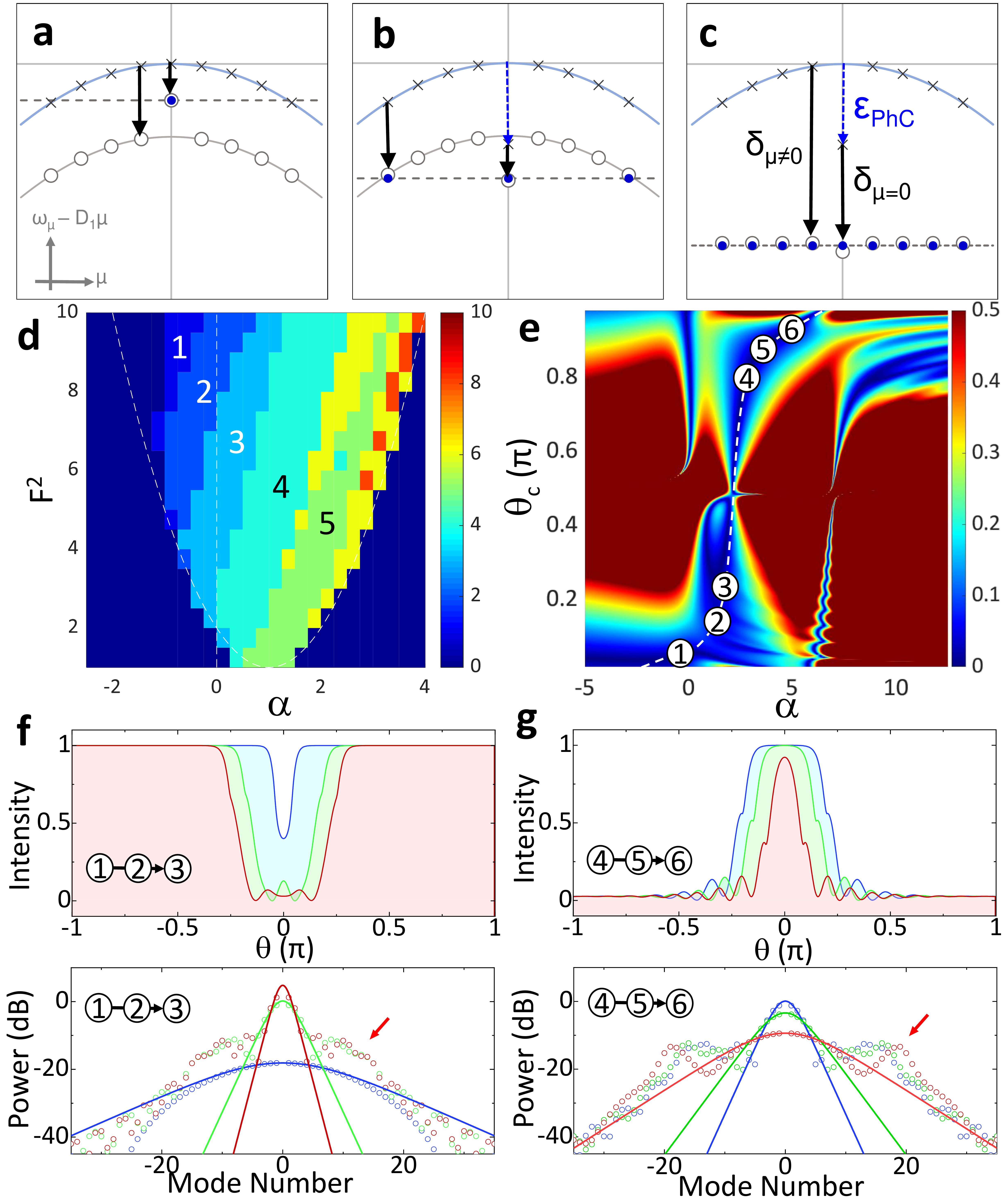} 
\caption{
	Kerr shift diagrams for
	\textbf{a} a normal dispersion base ring,
	\textbf{b} the Turing pattern and
	\textbf{c} the pulse state in the PhCR, showing the cold-cavity ($\times$) and Kerr-shifted ($\circ$) frequencies, and energy in some modes (\textcolor{blue}{$\bullet$}).
	\textbf{d} Diagram showing number of fringes in the optical waveform.
	\textbf{e} Error measure of the analytical Ansatz using Eq.~S\ref{m12Mismatch} (for $ F=3.0 $), where a low-error valley (dashed white line) indicates reducing filling fraction with detuning.
	\textbf{f-g} show the intensities and spectra of the Ansatz along the curve in \textbf{e}, sweeping from \textbf{f} dark soliton toward half-filled, and
	\textbf{g} half-filled toward bright pulse.
	The center lobe fits on the spectra (plain lines) are $y=10 \log_{10}(\sech^2(\frac{x-x_0}{BW}))$.
}
\label{Fig:STheory} 
\end{figure}

Phase-matching for pattern generation is conventionally unavailable in normal dispersion resonators. In this work, this is overcome using the photonic crystal shift. Figure~S\ref{Fig:STheory}\textbf{a} illustrates the integrated dispersion $D_\text{int}(\mu) = \omega_\mu - \omega_0 - \mu D_1$, where $\omega_\mu$ stands for the cold-cavity frequencies for the $\mu$-th mode from the pump mode $\mu=0$, $\omega_0$ the pump mode frequency, and $D_1$ the local mode spacing or free-spectral range. As the $\mu=0$ mode is energized by a pump laser, optical intensity builds up in the resonator, causing the modes to shift toward lower frequencies under the Kerr effect. However, the pump mode self-frequency shift  $\delta_{\mu=0} = \frac{1}{2} \delta{\mu'\neq 0}$ is a factor of two smaller than the other modes experiencing cross phase modulation~\cite{FactorOfTwoBook}. In an anomalous dispersion system, the $D_\text{int}$ curve has positive curvature, meaning the frequency difference from the local mode spacing increases with $\mu'$. This compensates for the Kerr shift mismatch between the $\mu=0$ mode and the $\mu'\neq 0$ modes, therefore enabling four-wave mixing (FWM) phase-matching. In the normal dispersion case shown in Fig.~S\ref{Fig:STheory}\textbf{a}, the curvature is negative, moving the $\mu'\neq 0$ modes further away from FWM phase-matching. Therefore, Kerr comb generation is absent in the conventional normal dispersion resonators.

A point-defect at the pump mode $\epsilon_\text{PhC}$ re-enables the FWM matching, shown in Fig.~S\ref{Fig:STheory}\textbf{b}, by filling in the mismatch between a desired pair of modes $\pm \mu'$ and the $\mu=0$ mode. The FWM matched modes can energize to form Turing patterns similar to the anomalous dispersion case. More importantly, the pump mode shift modifies the detuning ranges where the pulse patterns form \cite{Yu2020}, also visible in main text Eq. \eqref{Eq:PSLLE}. The stationary waveforms in the Kerr resonator like the bright- or dark-pulse states compose of many interlocking modes. Their component mode frequencies pull into alignment with each other by Kerr shift~\cite{Bao2014}, shown in Fig.~S\ref{Fig:STheory}\textbf{c}. Since the self- and cross-phase modulation difference does not depend on the sign of dispersion, the counter-balancing term in Eq. \eqref{Eq:PSLLE} can also be interpreted as the time-domain equivalent of the Kerr-mismatch balancing in Ref.~\cite{Yu2020}.

\section{Derivation of Local Energy Flow} \label{S:RhoDerive}

We provide the derivation for Main Text Eq.\eqref{Eq:EnergyFlow} by evaluating the rate of change for the intensity $I(\theta)$, which is the energy per unit $\theta$. We calculate this quantity by substituting Main Text Eq. \eqref{Eq:PSLLE} into the expression $\partial_t I(\theta)$:
\begin{align*}
	\partial_t I =& \psi^* \partial_t \psi + h.c. \\
	=& -2|\psi|^2 - \frac{i\beta}{2} ( \psi^* \partial_\theta^2\psi - \psi \partial_\theta^2\psi^* )\\
	&+ F(\psi^*+\psi) + i\epsilon (\psi^*\Bar{\psi} - \psi\Bar{\psi}^*)\\
	=& -2 I +\frac{\beta}{2} \cdot 2 \partial_\theta \cdot \mathbb{I}m(\psi^*\partial_\theta\psi)\\
	&+ 2 F \cdot \mathbb{R}e(\psi) +2 \epsilon \cdot \mathbb{I}m(\psi \Bar{\psi}^*)\\
	=& 2 \Big( -\rho_\text{loss} - \triangledown_\theta\cdot J_\beta + \rho_\text{in} + P_\text{PhC} \Big)
\end{align*}%
\noindent where $\rho_\text{in}$, $\rho_\text{loss}$ are the energy in-flow and loss defined in the Main Text, $J_\beta = -\frac{\beta}{2}\cdot \mathbb{I}m(\psi^*\partial_\theta\psi)$ is an energy current driven by dispersion, and $P_\text{PhC} = \epsilon \cdot \mathbb{I}m(\psi \Bar{\psi}^*)$ represents power exchange induced by the photonic shift $\epsilon$. We identify $\rho_\text{in}$ as the energy flow of the physical pump as it is proportional to $F$. For the discussion of the two intensity levels, the local field is approximately flat, $J_\beta \simeq 0$. We also note that $P_\text{PhC}$ exchanges energy within the resonator, but $\oint P_\text{PhC} \cdot d\theta = \mathbb{I}m (|\Bar{\psi}|^2)=0$, conserving the total energy.

\section{Bandwidth to filling-fraction correspondence} \label{S:BWtcDerive}

We derive the inverse relation between fitted bandwidth $BW$ in main text Fig.~\ref{Fig:Spectra} and the filling fraction parameter $t_c$ in the theoretical analysis. We begin with the square-wave function:

\begin{equation*}
    \psi(\theta) = \Theta(\pi t_c -|\theta|)
\end{equation*}

\noindent where $\Theta(\theta)$ is the Heaviside function. We obtain the modal spectrum of this waveform by carrying out the Fourier transform:

\begin{equation*}
v(m) = \hat{\mathcal{F}}(\psi) = \frac{sin(m\pi t_c)}{m\pi}
\end{equation*}

\noindent where we note that setting $t_c \rightarrow 1-t_c$ changes the modal phase by $(-1)^{m+1}$ but leaves the magnitude invariant, as a result of the bright- and dark-pulse correspondence.

We focus on the center-lobe near $m=0$ and calculate the second-order derivative which links to the bandwidth. Specifically, we compute the second-order derivative for the spectral power in units of dB, $S(f) = 10 log_{10}(|v(f)|^2)$, where the mode frequencies are $f = c + m\cdot \text{FSR}$, where FSR is the free-spectral range. Evaluating the second derivative of $S$ at $m=0$, we get:

\begin{equation*}
\partial^2_f S(f=c) = -\frac{20}{ln(10)} \frac{\pi^2 t_c^2}{3 \text{FSR}^2}
\end{equation*}

\noindent where $ln(x)$ is the natural-log function. We compare this to the fitting function to extract BW:

\begin{equation*}
y = a + 10 \log_{10}\left[ \sech^2((f-c)/{\rm BW}) \right]    
\end{equation*}

\noindent by taking its second-order derivative at $f=c$, we get:

\begin{equation*}
\partial^2_f y(f=c) = -\frac{20}{ln(10)}\cdot\frac{1}{\text{BW}^2}
\end{equation*}

\noindent Comparing the two forms, we obtain the relation between BW and $t_c$:

\begin{equation*}
\frac{1}{\text{BW}^2} = \frac{\pi^2 t_c^2}{3 \text{FSR}^2}
\end{equation*}

\noindent taking the square-root of both sides, we get the relation:

\begin{equation*}
\text{BW} = \frac{\sqrt{3} \text{FSR}}{\pi}\cdot t_c^{-1}
\end{equation*}

\noindent which shows the inverse-proportional relation between $BW$ and $t_c$, or, in the case of the dark-pulse-like ranges:

\begin{equation*}
\text{BW} = \frac{\sqrt{3} \text{FSR}}{\pi}\cdot (1-t_c)^{-1}
\end{equation*}

\noindent In summary, the spectral manifestation of varying temporal filling fraction with laser detuning is an inverse-proportional change in the center-lobe bandwidth.


\section{Waveform Features}

The waveforms in normal dispersion systems show several distinct features. The main text focuses on the center lobe and its implication on the temporal duration of the bright- or dark-pulse. There are additional features at higher azimuthal frequencies including the `wing' and `horn'. The wing feature arises from the dark pulse touching down to zero-intensity and developing internal features (red arrow in Fig.~S\ref{Fig:STheory}\textbf{f}). The horn feature comes from the oscillating patterns near the body of the bright-pulse (red arrow in Fig.~S\ref{Fig:STheory}\textbf{g}). The phase diagram of the developing of these patterns, shown by plotting the number of local minima in the field, is shown in Fig. S\ref{Fig:STheory}\textbf{d}. We demonstrate the origin of these features using a locally linearised LLE with a piecewise-constant intensity nonlinear term. This allows us to derive an approximate waveform solution for given system parameters (detuning $\alpha$ and angle $\theta_c$ separating the two piecewise domains). The panels Fig.~S\ref{Fig:STheory}\textbf{f},\textbf{g} are created using this method and reasonably approximate the full LLE waveforms.

We write the LLE in the following form:
\begin{equation} \label{LocalLLE}
    \partial_\tau \psi = -(1+i \alpha)\psi -\frac{i d_2}{2}\partial^2_\theta \psi + i I(\theta) \psi + F
\end{equation}
\noindent where $I(\theta)$ is the local intensity at azimuthal angle $\theta$. We approximate the equation by separating the resonator into the domains $|\theta| < \theta_c$ and $|\theta| > \theta_c$, where $\theta_c$ specifies a switching azimuthal angle between the two domains, related to the filling ratio by $\theta_c = (1-t_c)~\pi$. We then assume $I(\theta)$ can be treated at a constant for each domain, switching between two fixed levels $I_j$, $j=1,2$. This method is reminiscent of the switching waves~\cite{Coen1999}, but is subjected to the edge-less boundary conditions of the ring resonator. The particular solution to the system are constant fields sourced by the pump field $F$ that produce the intensities $I_j$ in a self-consistent manner. We then search for the general solutions to Eq.~S\ref{LocalLLE} under this approximation. The linearized second-order differential equation in $\theta$ reads:
\begin{equation}
\frac{i \beta}{2}\partial^2_\theta \psi = -(1+i \alpha_j)\psi
\end{equation}
\noindent where $\alpha_j=\alpha-I_j,\ j= 1,2$ are constants for each domain. The equation yields general solutions of the form $\exp(\lambda_\pm \theta)$, where \(\lambda_\pm = \pm \frac{1}{\sqrt{\beta}} (u(\alpha_j)+i u(-\alpha_j))\), \(u(x)=\sqrt{\sqrt{x^2+1}-x}\). We choose solution functions with the form:
\begin{equation} \label{Ansatz}
    \psi_j(\theta)=A_j \cosh(\lambda_j \theta) + E_j
\end{equation}
\noindent for each domain, where $E_j$ is the background field for level $j$, $I_j = |E_j|^2$, and the hyperbolic cosine function is selected to respect the symmetry in $\pm\theta$. The pattern of the complex hyperbolic cosine can be exponential-like or sine-like depending on the signs of the local $\alpha_j$. Finally, we set the the background level $E_1$ for $|\theta|<\theta_c$, and $E_2$ for $|\theta|>\theta_c$. We solve for the coefficients $A_j$ by requiring field continuity $\psi_1(\theta_c)=\psi_2(\theta_c)$, and continuity of the derivatives $\psi_1'(\theta_c)=\psi_2'(\theta_c)$. This process provides us with an analytical Ansatz for the field in the resonator, for each parameter set $(\beta, \alpha, \theta_c, E_j)$. We find this simple approximate solution is sufficient to reproduce the spectral features we observed.

We identify the Ansatz parameters best matching the physical state by minimizing their error in the time-stationary LLE. In particular, we search for a fitness measure that is sensitive to the intensity filling fraction by examining the role of each mode order $\mu$. The pump mode $\mu = 0$ creates a flat background, which interferes with mode $\mu' = \pm 1$ to create a simple cosine modulation. At this point, the filling fraction is exactly 0.5 due to the shape of the cosine function. Adding $\mu' = \pm 2$ terms, we begin to modify the filling ratio, depending on the relative phase between the modes. For example, $1+0.75~cos(\theta) + 0.25~cos(2\theta)$ creates a bright pulse, while $1+0.75~cos(\theta) - 0.25~cos(2\theta)$ creates a dark pulse. With this in mind, we create a fitness function depend on the $|\mu|\leq 2$ modes. We create a simple form by dividing the modal equations for mode 1 and 2 by their respective field amplitudes, and subtracting the two. This creates an advantageous form that eliminates explicit dependence on the pump mode shift $\epsilon_\text{PhC}$ or the pump field parameters $\alpha$, $F$:
\begin{equation} \label{m12Mismatch}
\xi_{12} = | \left(\delta_2 -d_\text{int}(2)\right) - \left(\delta_1-d_\text{int}(1)\right) |
\end{equation}

\noindent where $\delta_j$, j=1,2 are the modal Kerr shift \cite{Yu2020}:

\begin{equation} \label{KerrShift}
\delta_\mu=\Re\left(\hat{ \mathcal{F}} \{|\psi(\theta)|^2\psi(\theta)\},_\mu / \hat{ \mathcal{F}} \{\psi(\theta)\},_\mu \right)
\end{equation}

\noindent and $d_\text{int}$ are the linewidth-normalized integrated dispersion. The $\xi_{12}$ term vanishes for time-stationary solutions of the LLE. We use this metric to minimize the error of our Ansatz. Figure~S\ref{Fig:STheory}\textbf{e} shows the evolution of $\xi_{12}$ as a function of $(\alpha,\theta_c)$, where we have assumed the two intensity levels $I_{1,2}$ correspond to the upper- and lower-state of the bi-stable pump mode on resonance~\cite{Godey2014} for $F=3.0$. $\xi_{12}$ shows three minimum valleys. The center valley (dashed line in Fig.~S\ref{Fig:STheory}\textbf{e}) shows the correct exponential-like shape for the upper level and sine-like shape for the lower, in agreement to calculated intensity patterns in the LLE. Tracing this minimum valley of $\xi_{12}$, we observe the trend of increasing $\alpha$ resulting in decreasing $t_c$ (increasing $\theta_c$). This suggests that the Ansatz functions, while crude, seem to capture important aspects of the system. The intensity patterns and spectra along the curve are shown in Figure~S\ref{Fig:STheory}\textbf{f,g}, at point 1--3 for the dark pulse, and 4--6 for the bright pulse.

The Ansatz functions show the shared physical origin of the lobe-number inside a dark-pulse ( Fig. S\ref{Fig:STheory}\textbf{f}) and oscillations on the sides of the bright-pulse ( Fig. S\ref{Fig:STheory}\textbf{f}). These two appear spectrally as the `wing' and `horn' features, respectively. The oscillations result from the sine-like behaviour of the linearized LLE in the low-intensity domain, while the periodicity approximately scales with $\sqrt{\beta}$ from the form of the eigenvalue $\lambda_\pm$. The upper level of the pulse has exponential-like waveform, therefore does not show the oscillations.

\section{Power Conversion Efficiency}

The 25\% theoretical limitation to power conversion efficiency results from an interplay between pump mode coupling condition and energy distribution of the Kerr effect. The Kerr effect extract energy from the pump mode and distributes it to the comb modes, resulting in an additional loss term on the $\mu=0$ mode, corresponding to the imaginary counterpart of Eq.~S\ref{KerrShift}. The pump mode coupling condition is affected by the Kerr term, in the form of a modified effective loss rate $\kappa^\text{pump}_i = \kappa_i + \kappa_\text{Kerr}$, where $\kappa_i$, $\kappa_\text{Kerr}$ stand for the intrinsic and Kerr-induced loss in the resonator. In the critical or near-critically coupled devices in this work, the comb formation becomes a self-limiting process. Increasing the pump power, thus increasing the strength of the Kerr-induced loss term, shifts the coupling condition toward under-coupling, preventing the pump power from entering the resonator. We study the optimal power-conversion pump $F^2$ in the LLE to find $\kappa^\text{pump}_i = \kappa_i + \kappa_c$ at the case giving the 25\% efficiency, an analogue condition to maximizing the in-cavity intensity in a Fabry-Perot resonator.

The total power efficiency $\eta_\text{total} = P^\text{out}_\text{comb}/P^\text{in}_\text{pump}$ can be improved by overcoupling the resonator. We specify the coupling condition by defining the coupling constant $K = \kappa_c / \kappa_i$. The total power efficiency can be related to the internal efficiency $\eta$ in the form:
\begin{equation} \label{EtaTtl}
\eta_\text{total} = \dfrac{4 K}{K+K^{-1}+2} \cdot \eta = \left(\dfrac{2K}{K+1} \right)^2 \cdot \eta
\end{equation}

\noindent where $K>1$ indicates over-coupling. The improved efficiency is achieved at the cost of increased threshold power, therefore reducing the maximum $F^2$ values that can be achieved by a given pump laser. The benefit of overcoupling arrives from that the absorption loss of the resonator is diluted by the rapid removal of energy by the bus waveguide, thus enabling us to overcome the 25\% energy division limit. Note that in the limit of high coupling $K \gg 1$, $\eta_\text{total} \rightarrow 4 \eta$, therefore the maximum internal efficiency of 25\% corresponds to an 100\% total efficiency in the limiting case. We are actively exploring stronger coupling parameter spaces, and also methods to control the pump mode coupling strength deferentially from the other modes, such as modified bus waveguides with a low-finesse pump recycling cavity~\cite{Sato2000} or contra-directional grating coupler~\cite{Shi2013}, to further improve the conversion efficiency.

\end{document}

%% file: preamble.tex
\usepackage{amsthm}
\usepackage{mathtools}
\usepackage{physics}
\usepackage{xcolor}
\usepackage{graphicx}
\usepackage[left=23mm,right=13mm,top=35mm,columnsep=15pt]{geometry} 
\usepackage{adjustbox}
\usepackage{placeins}
\usepackage[T1]{fontenc}
\usepackage{lipsum}
\usepackage{csquotes}
\usepackage[version=3]{mhchem}
\usepackage[squaren, Gray, mediumqspace]{SIunits}